\documentclass[a4paper,fleqn]{cas-dc}

\usepackage[numbers]{natbib}

\usepackage{amssymb}
\usepackage{amsmath}
\usepackage{chemformula}
\usepackage[capitalize]{cleveref}
\usepackage{xspace}
\usepackage{booktabs}

\newcommand{\MAX}{\ch{(Cr$_{0.5}$Mn$_{0.5}$)$_2$GaC}\xspace}
\newcommand{\MAXgeneral}{\ch{(Cr$_{1-x}$Mn$_x$)$_2$GaC}\xspace}

\def\tsc#1{\csdef{#1}{\textsc{\lowercase{#1}}\xspace}}
\tsc{WGM}
\tsc{QE}
\tsc{EP}
\tsc{PMS}
\tsc{BEC}
\tsc{DE}

\begin{document}
\let\WriteBookmarks\relax
\def\floatpagepagefraction{1}
\def\textpagefraction{.001}
\shorttitle{Laser-inducd demagnetization in a MAX phase}
\shortauthors{Mogunov~Ia.~A.~et~al.}

\title[mode=title]{Laser-induced demagnetization in a MAX phase (Cr$_{0.5}$Mn$_{0.5}$)$_2$GaC}

\author[1]{Iaroslav Mogunov}[orcid=0000-0002-7650-0074]
\cormark[1]
\cortext[cor1]{Corresponding author.}
\ead{mogunov@mail.ioffe.ru}
\credit{Conceptualization,
Data curation,
Formal analysis,
Funding acquisition,
Investigation,
Project administration,
Validation,
Writing – original draft,
Writing – review \& editing}

\author[1]{Artyom Gorshkov}
\credit{Data curation,
Formal analysis,
Investigation,
Methodology,
Software,
Visualization}

\author[2]{Mikhail Rautskii}
\credit{Formal analysis,
Investigation}

\author[2]{Tatyana Andryushchenko}
\credit{Resources,
Writing – review \& editing}

\author[1]{Alexandra Kalashnikova}
\credit{Conceptualization,
Formal analysis,
Methodology,
Supervision,
Writing – original draft,
Writing – review \& editing}

\affiliation[1]{organization={Ioffe Institute},
                addressline={Politekhnicheskaya 26}, 
                city={St. Petersburg},
                postcode={194021}, 
                country={Russia}}
\affiliation[2]{organization={Kirensky Institute of Physics, Federal Research Center KSC SB RAS},
                addressline={Akademgorodok 50, bld. 38}, 
                city={Krasnoyarsk},
                postcode={660036}, 
                country={Russia}}

\begin{abstract}
Magnetic MAX phases are nanolaminated metals that combine ceramic-like thermal and mechanical stability with peculiar magnetic ordering, making them attractive for thin-film optoelectronics and spintronics.
However, laser-induced magnetization dynamics in MAX phases remains largely unexplored.
Here, we investigate laser-induced ultrafast demagnetization in a 40-nm-thick epitaxial film of the magnetic MAX phase \MAX, which magnetically orders below $\sim$220~K, using time-resolved magneto-optical Kerr effect spectroscopy.
We reveal, that the demagnetization transients exhibit a two-step type-II demagnetization, common for two-dimensional magnetic systems.
The second stage dominates the process and has a characteristic time of approximately 100~ps.
Applying the three-temperature model, we extract the electron-lattice, spin-lattice, and electron-spin coupling constants.
The reconstructed spin heat capacity exhibits a weak temperature dependence, accounting for only a subtle slowing down of demagnetization at elevated temperatures and fluences.
\end{abstract}

\begin{keywords}
Ultrafast demagnetization 
\sep MAX phase
\sep 2D magnets
\sep 3-temperature model
\end{keywords}

\maketitle

\section{Introduction}
\label{sec:intro}

The rise of interest in two-dimentional (2D) materials was initiated by the fabrication of graphene \cite{Geim_science_review_2009} and has since led to the demonstration of numerous unique properties in a diverse family of 2D structures \cite{Ren_2D_review_2026}.
Nowadays, these materials are regarded as a promising multifunctional platform for optoelectronics and spintronics \cite{Tan_2d_optoelecronics_review_2020,Ahn_2D_spintronics_review_2020,Sierra_2D_spintronics_review_2021}.
Compounds with the formula M$_{n+1}$AX$_n$ (where M is a transition metal, A is an A-group element, X is C or N, and $n=1..4$), known as MAX phases, consist of M$_{n+1}$X$_n$ layers linked by a layer of A ions.
This family of materials uniquely combines the thermal and mechanical resilience of ceramics with metallic conductivity \cite{zhang-NanoMicro2025}.
The exceptionally wide range of synthesized MAX phases \cite{dahlqvist-MaterToday2024,zhang-NanoMicro2025}, and a possibility of extracting 2D M$_{n+1}$X$_n$ layers, MXens \cite{Anasori-2D2022}, position them as strong candidates for constructing versatile 2D heterostructures.

Motivated by the general interest in atomically thin magnetically ordered materials \cite{lee-Nano2016,huang-Nature2017}, significant research efforts are focused on the fabrication and exploration of magnetic MAX phases as nanolaminated magnets.
Theoretically, a variety of spin arrangements can be realized in these compounds due to competing intra- and interlayer exchange couplings \cite{Ingason-JPhysCM2016,Dey-PRBPhysRev2023,Tao_rare-earth_frustrated_mag_iMAXes_2019,Hamm_mag_MAX_on_A_pos_2017}.
Consequently, establishing the exact magnetic structure of MAX phases experimentally has proven to be a challenging task \cite{Dey-PRBPhysRev2023,Dahlqvist-PRB2016,Liu-PRB2013,Thorsteinsson-PRMater2023,kubitza-ChemMat2024}.

Ultrafast lasers have lead to the development of novel experimental techniques and the advancement of existing ones, with far-reaching applications across various scientific disciplines.
In magnetism, it has enabled the study of magnetization dynamics on timescales ranging from attoseconds \cite{siegrist-Nature2019} to milliseconds \cite{Wang-PRRes2021} and even seconds \cite{Medapalli-PRB2017}, thereby promoting a unified and hierarchal view on the coupling of spins to the other energy and angular momentum reservoirs.
Ultrafast laser-induced demagnetization, or a rapid partial loss of magnetization followed by a slower recovery, has been shown to be highly sensitive to the electronic band structure, crystal and magnetic structures and their corresponding excitations, heat capacities, and mutual couplings \cite{Chen-PhysRep2025}.
Striking examples include the two distinct types of demagnetization observed in transition and rare-earth metals \cite{Koopmans-NMater2010}, the critical slowing down of demagnetization near the Curie temperature \cite{lopez2013-PRB2013,Kimling-PRB2014,Roth-PRX2012}, and the generally slower demagnetization in dielectrics \cite{kimel-PRL2002}.
Naturally, many features of ultrafast demagnetization become particularly pronounced in magnetic materials with reduced dimensionality, as demonstrated for van der Waals (vdW) materials \cite{Zhang-NanoLett2021,Lichtenberg-2D2023,Kuntu-PRMater2024,Wang-NSR2025,Sun_vdW_demag_2021,Wu_FeGeTe_2024,Tomarchio_FeGeTe_2024,Sutcliffe_PRB_CrGeTe_2023,Khusyainov_CoPS3_2023} and Kagome-lattice antiferromagnets \cite{kong-ACS2025}.
However, ultrafast demagnetization in MAX phases remains largely unexplored, raising the question of whether it follows trends similar to those observed in other layered magnets.

In this Article, we employ ultrafast demagnetization as a tool to elucidate the magnetic properties of the MAX phase \MAX, which have so far been only partially characterized.
\MAXgeneral family is known for its relatively high magnetic ordering temperatures that can be controlled by the composition $x$.
We find that a 40-nm-thick film of \MAX exhibits type-II demagnetization behavior across the studied ranges of laser fluence and temperature.
Using a three-temperature model, we estimate the electron-lattice, spin-lattice, and electron-spin coupling constants, which are then compared to those of van der Waals magnets.
Furthermore, we quantify the spin-specific heat and reveal its evolution as the temperature approaches the magnetic order-disorder phase transition temperature $T_C$.

\section{Experimental}
\label{sec:experimental}

\subsection{MAX phases \MAXgeneral and \MAX sample}

\begin{figure}
    \centering
    \includegraphics[width=\columnwidth]{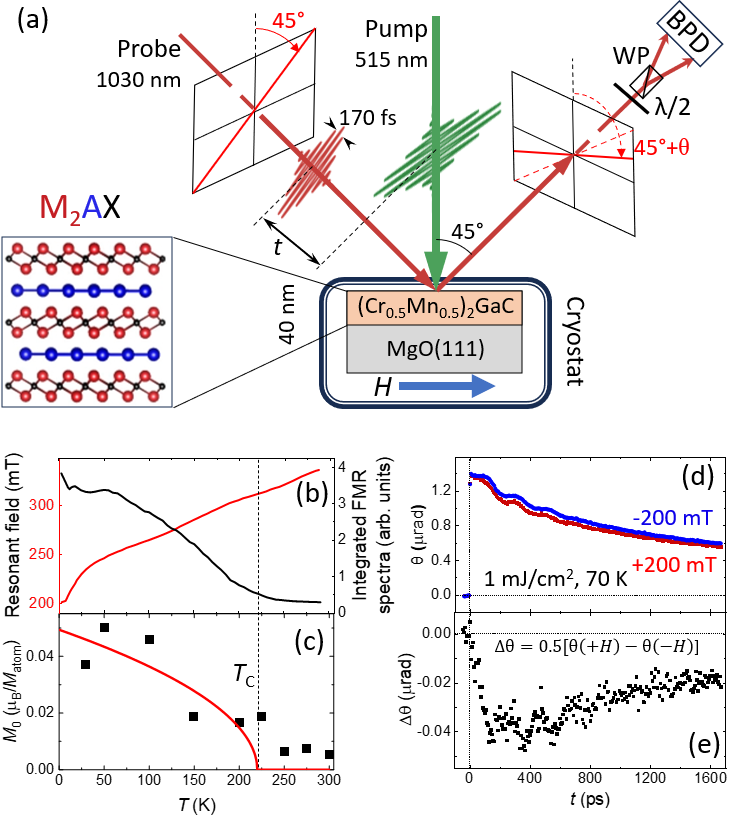}
    \caption{       
      (a)~Schematic of the pump-probe time-resolved magneto-optical Kerr effect (tr-MOKE) setup, illustrating the experimental geometry.
      BPD -- balanced photodetector, $\lambda$/2 -- half-wave plate, WP -- Wollastone prism.
       The inset shows the crystal structure of \MAXgeneral;
      (b)~Temperature dependence of the ferromagnetic resonance (FMR) integral (black, right) and the corresponding resonant field (red, left);
      (c)~Remanent magnetization $M_0$ taken from Ref.~\cite{Petruhins-JMatSci2015} (dots) and fit by the Curie-Weiss law (line);
      (d)~Representative polarization rotation signals $\theta(t)$ measured at fixed temperature and pump fluence for two opposite directions of the applied magnetic field;
      (e)~Difference between the signals shown in (d), revealing the magnetic-field-odd ($H$-odd) contribution to the measured response.
    }\label{fig:setup}
 \end{figure}

\MAXgeneral crystallizes in a hexagonal structure (space group $P6_{3}/mmc$).
The structure consists of (Cr$_{1-x}$Mn$_x$)$_2$C layers stacked along the hexagonal axis and separated by Ga planes, as shown schematically in Fig.~\ref{fig:setup}(a).
Bonding is covalent within the (Cr$_{1-x}$Mn$_x$)$_2$C layers and metallic between these layers and the Ga sheets \cite{Thorsteinsson_MnCrGaC_system_2023}.
The in-plane lattice parameter increases with $x$ laying between $a$=2.89-2.91~\AA, while the out-of plane parameter decreases with $x$ and lays between $c$=12.55-12.62~\AA~\cite{Thorsteinsson_MnCrGaC_system_2023,Siebert_CrMnGaC_system_2021,Petruhins-JMatSci2015}.

The most detailed theoretical and experimental investigations of magnetism in the \MAXgeneral family have been reported for the compound with $x=1$, which exhibits a magnetic ordering at $T_C$ above room temperature \cite{Dey-PRBPhysRev2023,Dahlqvist-PRB2016,Ingason-PRB2016}.
It is established that there is a robust intralayer ferromagnetic coupling \cite{Dahlqvist-PRB2016,Ingason-PRB2016}.
Considering the magnetic moments within each layer as a single "supermoment", the overall magnetic structure below $T_C$ is an incommensurate spiral formed by these macrospins \cite{Dey-PRBPhysRev2023}.
This structure results in a specific magnetic field dependence of magnetization, characterized by low remanence and high saturation fields \cite{Dahlqvist-PRB2016,Thorsteinsson_MnCrGaC_system_2023}.
Additionally, at 250~K a transformation from an antiferromagnetic to a canted antiferromagnetic structure occurs, concomitant with a reduction of the interlayer distance \cite{Dahlqvist-PRB2016,novoselova-SciRep2018}.
Owing to its nanolaminated nature, the compound possesses easy-plane magnetic anisotropy, with magnetic moments aligned within the Mn$_2$C layers \cite{Thorsteinsson-PRMater2023}.

Decreasing $x$ reduces $T_C$ \cite{Thorsteinsson_MnCrGaC_system_2023,Thorsteinsson_MnCrGaC_system_2025,Yan_CrMnGaC_system_2020} and modifies the magnetization.
The compound with $x$=0.5 was reported to exhibit the highest magnetization among Mn-containing MAX phases \cite{Petruhins-JMatSci2015}, later challenged by Ref.~\cite{Thorsteinsson_MnCrGaC_system_2025}.
For \MAX, a magnetization behavior analogous to that of Mn$_2$GaC was reported \cite{Petruhins-JMatSci2015}, suggesting noncollinear antiferromagnetic ordering of the "supermoments".
Based on magnetometry and ferromagnetic resonance measurements, long-range order occurs below $T_C$, which lies in the range of 205-250~K \cite{Petruhins-JMatSci2015,Salikhov-MatResLett2015,Lai-APLMater2018}.
No vast evidence of an additional magnetic phase transition below $T_C$ has been reported for \MAX.

In this study, we used a 40-nm-thick epitaxial film with composition \MAX grown on a MgO(111) substrate by DC-magnetron sputtering from elemental sources C (99.99~\% purity) and Ga (99.9999~\% purity), and a compound Cr/Mn target with 50/50 at.\% (99.95~\% purity) \cite{Petruhins-JMatSci2015}.
The pressure of argon in the chamber (with base pressure below 5$\cdot$10$^{-9}$~Torr) during the
deposition was 4.5~mTorr \cite{Petruhins-JMatSci2015}.
X-ray diffraction analysis of the film indicated a single-phase epitaxial growth of \MAX with (000l) orientation of crystallites.
The lattice parameters of \MAX, modified by epitaxial strain, are: $a$=2.897\AA  (in-plane) and $c$=12.595\AA  (out-of-plane) \cite{Petruhins-JMatSci2015}.
The sample was stored at room temperature and relative humidity of 60$\pm$10~\% for four years.
Afterwards the sample was annealed in ultra-high vacuum (UHV) with a base pressure of 4$\cdot$10$^{-9}$~Torr at 300\textdegree C for 3 hours, etched by Ar ions (pressure of 10$^{-5}$~Torr) for 60 minutes, and annealed again in UHV at 300\textdegree C for 30 minutes to desorb argon.
The procedure lead to a 3~nm film thickness reduction (based on preliminary calibration).
The details of the sample post-proceeding, along with the results of Auger analysis of the film surface oxidation, are presented in Ref.~\cite{Andryushchenko_CrMnGaC_2023}.

The film was characterized by magnetometry in Refs.~\cite{Salikhov-MatResLett2015,Lai-APLMater2018} confirming magnetic ordering below $T_C$=220~K as well as easy-plane magnetic anisotropy.
Ferromagnetic resonance (FMR) measurements were performed using a frequency of 9.47~GHz with varying temperature confirming MAX phase stability an a longer time span as reported in earlier works on this sample \cite{Salikhov-MatResLett2015,Novoselova_CrMnGaC_2019}.
Figure~\ref{fig:setup}(b) shows the resulting temperature dependencies of resonant field (red line) and integrated FMR spectra (black line), the latter confirms the ordering temperature at $T_C$=220~K.

\subsection{Pump-probe setup}

Magnetic dynamics was investigated using a time-resolved magneto-optical Kerr effect setup (tr-MOKE), as schematically illustrated in Fig.~\ref{fig:setup}(a).
The laser source is a Yb:KGW femtosecond laser amplifier with central wavelength of 1030~nm, pulse duration of 170~fs and a repetition rate of 100~kHz.
The output is split into pump and probe beams, with a variable delay $t$ introduced by a mechanical delay line with an accuracy of 10~fs.
The pump beam is frequency-doubled using a beta-borate (BBO) crystal and incident normally onto the sample surface.
The probe beam is linearly polarized at 45~deg and is incident on the sample at an angle of 45~deg relative to the surface normal.
The elliptical spot sizes on the sample surface are 55~$\mu$m~$\times$~55~$\mu$m for the pump and 35~$\mu$m~$\times$~25~$\mu$m for the probe.
The penetration depths of the pump and probe beams, estimated from the optical properties of Mn$_2$GaC \cite{Lyaschenko_Mn2GaC_magopt_2021}, are 1/$\alpha_{515}$=17~nm for the pump and 1/$\alpha_{1030}$=24~nm for the probe, both values are smaller than the film thickness of 40~nm.

In experiments we detect the change in the polarization state $\theta(t)$ of the reflected probe beam using a half-wave plate, a Wollastone prism, and a balanced photodetector.
Lock-in detection is employed at the pump modulation frequency of 1~kHz.
The magnetic field up to $\mu_0H$=300~mT is applied both in the sample plane and in the plane of laser incidence, thereby realizing the longitudinal tr-MOKE geometry.
Consequently, the detected polarization rotation $\theta(t)$ is sensitive to transient changes in both the in-plane and out-of-plane magnetization components.
The sample is mounted in a cold-finger cryostat, allowing measurements in the temperature range of 70-300~K.
We note that the used range of field values is below $\mu_0H$=5~T required to saturate the sample \cite{Petruhins-JMatSci2015}.
The measured static-field dependences of MOKE confirmed that the magnetic structure is not saturated.
Nevertheless, a measurable tr-MOKE signal could be detected, as shown below.

\section{Results and discussion}
\label{sec:results_dicsussion}

\subsection{Experimental observation}

To isolate the magnetization-related contribution to $\theta(t)$, each measurement was performed for two opposite directions of the applied magnetic field, and the corresponding signals were subtracted from one another, as illustrated in Figs.~\ref{fig:setup}(d-e).
The resulting curves $\Delta\theta(t,|H|)=(\theta(t,H)-\theta(t,-H))/2$ contain only field-dependent contributions.
The $\Delta\theta(t)$ transients show an initial decrease followed by a recovery.
No consistent oscillatory behavior was observed; consequently, we attribute this $H$-odd contribution solely to laser-induced decrease of tr-MOKE signal stemming from the decrease of magnetization and hereafter refer to it as demagnetization.

\begin{figure}
    \centering
    \includegraphics[width=1\columnwidth]{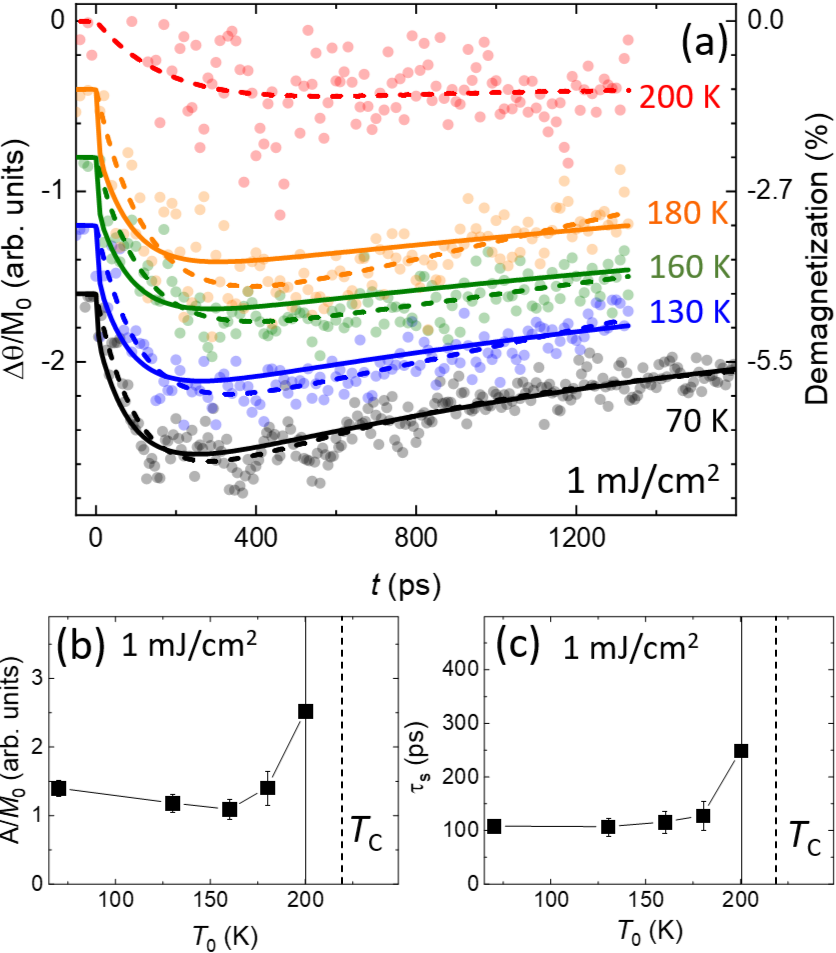}
    \caption{
        (a)~Magnetic-field-odd ($H$-odd) contribution to the measured signals for a laser fluence $F$=1~mJ/cm$^2$ and a range of temperatures (symbols).
        The data are vertically offset for clarity.
        Dashed lines represent fits using a three-exponential function, and solid lines show the results of the three-temperature model (3TM).
        Right-hand scale denotes the degree of demagnetization derived from the 3TM calculations;
        (b)~Temperature dependence (symbols) of the normalized demagnetization amplitude $A/M_0$ and (c)~time $\tau_s$, both obtained at $F$=1~mJ/cm$^2$.
        Lines are guides for the eye.
    }\label{fig:temperatures}
\end{figure}

\begin{figure}
    \centering
    \includegraphics[width=1\columnwidth]{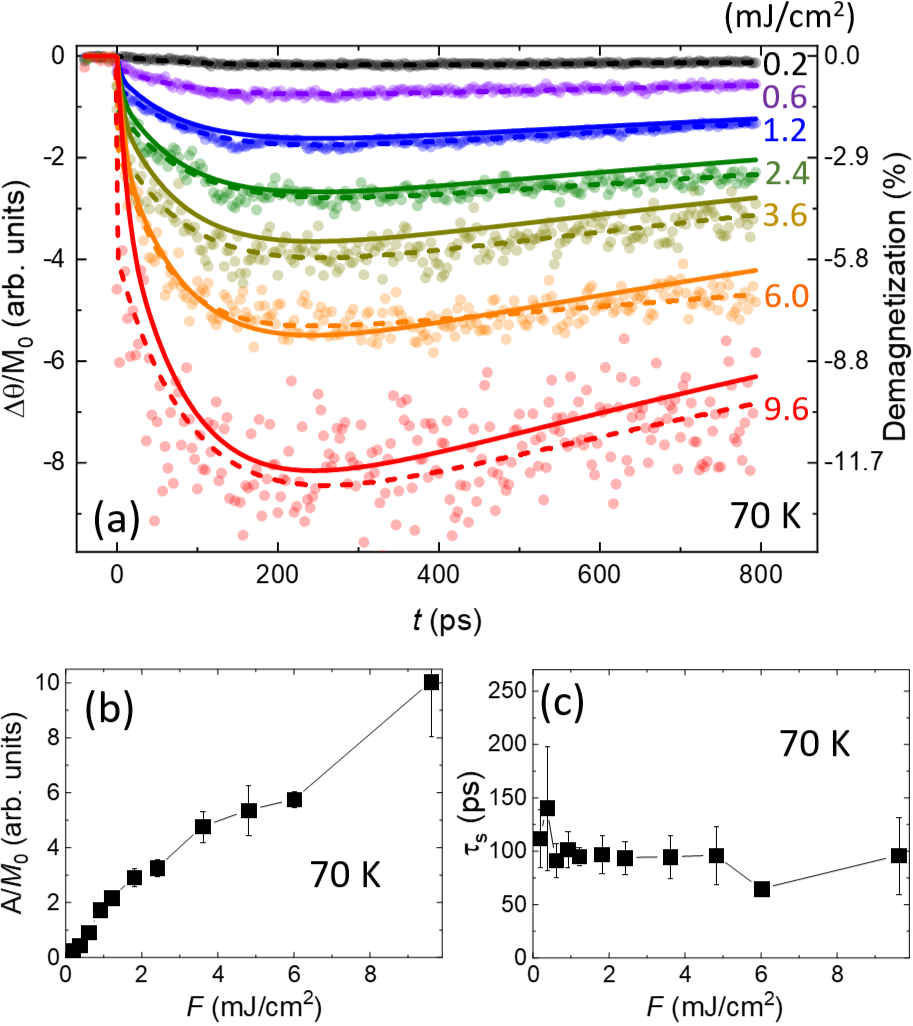}
    \caption{
        (a)~Magnetic-field-odd ($H$-odd) contribution to the measured signals for an initial temperature $T_0$=70~K and a range of laser fluences (symbols).
        Dashed lines represent fits using a three-exponential function, and solid lines show the results of the three-temperature model (3TM).
        Right-hand scale denotes the degree of demagnetization derived from the 3TM calculations;
        (b)~Fluence dependence (symbols) of the normalized demagnetization amplitude $A/M_0$ and (c)~time $\tau_s$ at $T_0$=70~K.
        Lines are guides for the eye.
    }\label{fig:fluences}
\end{figure}

We performed measurements for a set of temperatures both below and above the magnetic transition temperature $T_C$=220~K at a fixed fluence $F$=1~mJ/cm$^2$, and for a set of fluences at a fixed temperature $T_0$=70~K.
The resulting demagnetization curves are shown in Fig.~\ref{fig:temperatures}(a) for the temperature dependence and in Fig.~\ref{fig:fluences}(a) for the fluence dependence.
To analyze the demagnetization dynamics, the $\Delta\theta(t)$ curves were fitted using a three-exponential function, which assumes fast and slow demagnetization stages followed by a recovery, and accounts for the finite duration of the pump and probe laser pulses:

\begin{equation} \label{eq:biexp_fit}
\begin{split}
    &\Delta\theta=\Bigl[\bigl(A_f(\text{e}^{-t/\tau_f}-1)+A_s(\text{e}^{-t/\tau_s}-1)+\\
    &+A_r(1-\text{e}^{-t/\tau_r})\bigr)\cdot h(t)\Bigr]*G(\sqrt{2}t_0,t); \\
    &G(t_0,t)=\frac{2\sqrt{\ln2}}{t_0\sqrt\pi}\exp\left[-\frac{4\ln2\:t^2}{t_0^2}\right].
\end{split}
\end{equation}

\noindent where $A_f$ and $A_s$ are the amplitudes of the fast and slow demagnetization stages, respectively, with $A=A_f+A_s$ denoting the total demagnetization amplitude, $\tau_f$ and $\tau_s$ are the corresponding time constants of the fast and slow demagnetization stages, $A_r$ and $\tau_r$ describe the recovery process, $h(t)$ is the Heaviside step function, the symbol "$*$" denotes convolution, and $t_0$=170~fs is the duration of the laser pulse.
For the temperature-dependent data shown in Fig.~\ref{fig:temperatures}(a) we set $A_f$=0.
To normalize that demagnetization amplitude we used a temperature dependence of the remanent magnetization $M_0$ reported for the same sample in Ref.~\cite{Petruhins-JMatSci2015} and fitted using the Curie-Weiss law with $T_C$=220~K, see Fig.~\ref{fig:setup}(c).
The resulting quantity $A/M_0$ is proportional to the demagnetization degree $\Delta M/M_0$.
We note that the three-exponential function in Eq.~(\ref{eq:biexp_fit}) does not describe the physics underlining the observed magnetization dynamics and is used to characterize the experimental curves.

The extracted values of $A/M_0$ and $\tau_s$ are plotted as functions of temperature in Figs.~\ref{fig:temperatures}(b) and \ref{fig:temperatures}(c), respectively, and as functions of pump fluence in Figs.~\ref{fig:fluences}(b) and \ref{fig:fluences}(c).
Several observations can be made based on the data.
First, a clear two-step demagnetization process is evident for higher fluences in Fig.~\ref{fig:fluences}(a), which is characteristic of type-II demagnetization \cite{Chen-PhysRep2025}.
In metallic magnets, type-II demagnetization is typically observed at high fluences and temperatures near $T_C$ \cite{Kuntu-JMMM2025,Roth-PRX2012,Koopmans-NMater2010}, as well as in $f$-element magnets \cite{Wietstruk_Gd_Tb_XMCD_PRL_2011}.
The long duration of the slower stage $\tau_s$$\sim$100~ps is commonly observed in two-dimensional materials such as vdW flakes \cite{Khusyainov_CoPS3_2023,Kuntu-PRMater2024,Strungaru-PRL2025,Sutcliffe_PRB_CrGeTe_2023,Sun_vdW_demag_2021}, including the metallic ferromagnet Fe$_3$GeTe$_2$ \cite{Lichtenberg-2D2023,Wang-NSR2025,Wu_FeGeTe_2024,Tomarchio_FeGeTe_2024}.

From Fig.~\ref{fig:fluences}(a) it is seen that the first fast demagnetization stage becomes more prominent with increasing fluence and temperature.
The demagnetization time $\tau_s$ shows a slight increase with temperature (Fig.~\ref{fig:temperatures}(c)) which we attribute to subtle critical slowing down of the demagnetization process near $T_C$ \cite{lopez2013-PRB2013,Kimling-PRB2014,Roth-PRX2012}.
The weak temperature dependence of $\tau_s(T_0)$ and the absence of a fluence dependence $\tau_s(F)$ (Fig.~\ref{fig:fluences}(c)) suggest a modest increase of the spin heat capacity $C_s$ as the system approaches the phase transition region \cite{Kimling-PRB2014} in \MAX, since the demagnetization time can be estimated as $\tau_s=C_s/G_{sl}$, where $G_{sl}$ is the spin-lattice coupling constant \cite{Kuntu-PRMater2024}.
The temperature dependence of the demagnetization amplitude $A/M_0$ (Fig.~\ref{fig:temperatures}(b)) is negligible within an experimental error.
The amplitude $A/M_0$ increases nonlinearly with fluence, which is characteristic of the values of $F$ nearing the total demagnetization.

\subsection{3-temperature model (3TM)}

To substantiate the proposed connection between the observed demagnetization behavior and properties of the MAX phase, we employed the widely used three-temperature model (3TM) \cite{Chen-PhysRep2025,Kuntu-JMMM2025}.
This model assumes instantaneous thermal relaxation within each subsystem and treats them using individual temperatures $T_e, T_l, T_s$ and heat capacities $C_e, C_l, C_s$ for the electrons, lattice and spins, respectively.
Thermal equilibration between subsystems is governed by coupling constants for electron-phonon $G_{el}$, spin-phonon $G_{sl}$ and spin-electron $G_{es}$ interactions, which are assumed to have negligible temperature dependence.
Energy is supplied to the system by the pump pulse, which is assumed to be absorbed exclusively by the electronic subsystem.
The time-dependent power distribution of the pump is given by $P(t)=F(1-R)(1-e^{-1})G(t_0,t)\alpha^{-1}_{515}$ \cite{Kuntu-JMMM2025}, where $R$=0.484 is the reflection coefficient of \MAX at the pump wavelength estimated from the optical parameters of Mn$_2$GaC \cite{Lyaschenko_Mn2GaC_magopt_2021}, and $G(t_0,t)$ is the Gaussian intensity profile of the pump pulse (see Eq.~(\ref{eq:biexp_fit})).
Energy relaxation occurs solely through the lattice and is governed by thermal diffusion, expressed as an exponential energy loss with a time constant $t_3$.
Ultimately, the 3TM can be written as a system of differential equations \cite{Chen-PhysRep2025,Kuntu-JMMM2025}:

\begin{equation} \label{eq:3TM}
\begin{split}
    &C_e\frac{dT_e}{dt}=-G_{el}(T_e-T_l)-G_{es}(T_e-T_s)+P(t), \\
    &C_l\frac{dT_l}{dt}=-G_{el}(T_l-T_e)-G_{sl}(T_l-T_s)-\frac{T_p-T_0}{t_3}, \\
    &C_s\frac{dT_s}{dt}=-G_{es}(T_s-T_e)-G_{sl}(T_s-T_l).
\end{split}
\end{equation}

The system of equations~(\ref{eq:3TM}) is solved numerically using the standard ODE45 scheme.
Parameter optimization is performed using a particle swarm optimization algorithm to search for the global minimum in parameter space \cite{Couceiro_particle_swarm_2016}.
The coupling constants are unknown for MAX phases and are therefore treated as free parameters of the 3TM.
The temperature dependence of the lattice heat capacity $C_l(T)$ was fitted using the Einstein model function \cite{Sutcliffe_PRB_CrGeTe_2023} (Eq.~(\ref{eq:Debye})) based on the data for Cr$_2$GaC from Refs.~\cite{Lin_CrMnGaC_system_2013,Tong_Cr2GaC_Cr2GaN_2019}:

\begin{equation} \label{eq:Debye}
    C_l\propto\Theta_{Ein}^2\frac{\exp\Theta_{Ein}}{(\exp\Theta_{Ein}-1)^2};\;\Theta_{Ein}=T_{Ein}/T.
\end{equation}

\begin{figure}
    \centering
    \includegraphics[width=1\columnwidth]{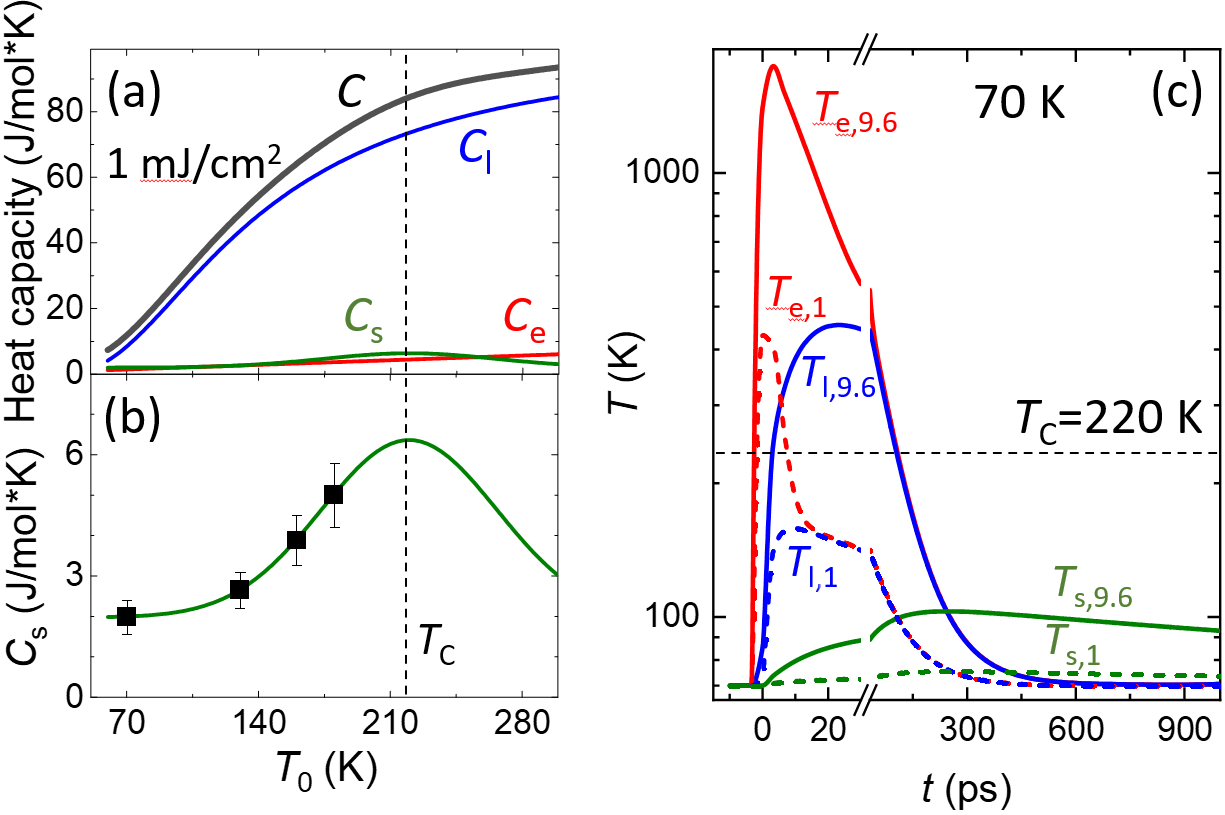}
    \caption{ 
        (a)~Temperature dependence of the calculated heat capacities: spin $C_s$ (green), phonon $C_l$ (blue), and electron $C_e$ (red), together with their sum $C$ (black), as obtained from the three-temperature model (3TM).
        (b)~A magnified view of $C_s(T)$, the symbols represent the values extracted from the 3TM, and the line is a Gaussian fit;
        (c)~Calculated temperature evolution of electrons $T_e$ (red), lattice $T_l$ (blue), and spins $T_s$ (green) at $T_0$=70~K for $F$=1~mJ/cm$^2$ (dashed lines) and $F$=9.6~mJ/cm$^2$ (solid lines).
        Note a logarithmic scale.
       }\label{fig:capacities}
\end{figure}

\noindent The used characteristic temperature was $T_{Ein}=0.75T_D=0.75\cdot570$~K \cite{Lin_CrMnGaC_system_2013}.
The resulting $C_l(T)$ is shown as the blue curve in Fig.~\ref{fig:capacities}(a).

The temperature dependence of the electronic heat capacity $C_e(T)$ was assumed to be linear, with a Sommerfeld constant $\gamma$=20.25~mJ/(mol K$^2$) \cite{Lin_CrMnGaC_system_2013}, and is shown as the red curve in Fig.~\ref{fig:capacities}(a).
The spin heat capacity $C_s$ was treated as constant for each value of $T_0$, allowing us to extract the temperature dependence of $C_s$ by applying the 3TM to the temperature-dependent demagnetization data shown in Fig.~\ref{fig:temperatures}(a).

The fitting of 3TM to the experimental data was based on the calculation of the spin temperature evolution $T_s(t)$.
We converted $T_s(t)$ into normalized magnetization and applied a convolution with the temporal shape of the probe pulse, as expressed in Eq.~(\ref{eq:biexp_fit}), using the relation:

\begin{equation}\label{eq:mag_from_Ts}
    1-\frac{\Delta M(t)}{M_0}=\left(1-\frac{M_0[T_S(t)]}{M_0[T_0]}\right)*G(t_0,t).
\end{equation}

\noindent The resulting $1-\Delta M(t)/M_0$ is proportional to the signals shown in Figs.~\ref{fig:temperatures}(a) and \ref{fig:fluences}(a), with the scaling factor being one of the fit parameters.
The solid lines in Figs.~\ref{fig:temperatures}(a) and \ref{fig:fluences}(a) represent $1-\Delta M(t)/M_0$ calculated using the 3TM.
The model successfully captures the measured demagnetization data.
The demagnetization is predominantly governed by a slow process, with signatures of a fast initial step appearing only at higher fluences and temperatures (see Fig.~\ref{fig:fluences}(a)).
However, the first fast stage was not well resolved experimentally, and therefore we do not discuss its parameters.

The resulting values of $C_s(T)$ are shown as symbols in the inset of Fig.~\ref{fig:capacities}(a).
At $T_0$=70~K, $C_s$=2~J/(mol K), which is in a reasonable agreement with the value of 1.25~J/(mol K) obtained as the difference between the total heat capacity $C$ of \MAX and that of non-magnetic Cr$_2$GaC reported in Ref.~\cite{Lin_CrMnGaC_system_2013}.
We then fit $C_s(T)$ with a Gaussian function centered at $T_C$, shown as the green line in Fig.~\ref{fig:capacities}(a) and \ref{fig:capacities}(b).
Typically, $C_s$ exhibits a sharp feature at $T_C$ \cite{Kormann_mag_heat_capacities_PRB_2011}, however, such a feature is expected to be smeared in thin films due to the presence of grains with different free energies, each affecting its local $T_C$ and hence broadening the transition for the film as a whole \cite{Salikhov-MatResLett2015}.
As a first approximation, this distribution can be modeled as a Gaussian, supporting our choice of function for $C_s(T)$.
The reconstructed $C_s(T)$ remains small up to $T_C$ and does not dominate the total heat capacity, unlike in other magnetic materials \cite{Kimling-PRB2014,Kormann_mag_heat_capacities_PRB_2011}, explaining the modest increase of the demagnetization time $\tau_s$ with temperature and fluence (Figs.~\ref{fig:temperatures}(c) and \ref{fig:fluences}(c)).

We fit both experimental datasets (Figs.~\ref{fig:temperatures}(a) and \ref{fig:fluences}(a)) with the same set of coupling parameters and the spin heat capacity depending only on the initial temperature $T_0$.
The extracted coupling constants are: $G_{el}$=(5.6$\pm$0.4)$\cdot$10$^{16}$, $G_{sl}$=(1.1$\pm$0.1)$\cdot$10$^{11}$ and $G_{es}$=(4.2$\pm$0.9)$\cdot$10$^{13}$~W/(m$^3$ K).
The electron-lattice coupling constant is similar to that in other vdW semiconductors \cite{Strungaru-PRL2025,Sutcliffe_PRB_CrGeTe_2023} and is considerably smaller than in a vdW metal \cite{Lichtenberg-2D2023,Adhikari_2025,Tomarchio_FeGeTe_2024}.
The electron-spin and spin-lattice coupling constants are several orders of magnitude smaller than in other vdW magnets \cite{Tomarchio_FeGeTe_2024,Khusyainov_CoPS3_2023,Kuntu-PRMater2024}.

Fig.~\ref{fig:capacities}(c) shows the evolution of the electronic (red), lattice (blue) and spin (green) temperatures following excitation by the laser pulse at $t$=0 with $F$=1~mJ/cm$^2$ (dashed lines) and 9.6~mJ/cm$^2$ (solid lines).
The resulting behavior is characteristic of a type-II demagnetization process \cite{Chen-PhysRep2025,Koopmans-NMater2010}, with $T_s$ lagging behind $T_l$.
For the used temperature and fluences, $T_s$ remains well below the disordering temperature $T_C$.
The modest heating of the spin subsystem is reflected in small values of the interaction constants $G_{es}$ and $G_{sl}$.

\section{Conclusions}

In conclusion, we have investigated ultrafast laser-induced demagnetization in a 40-nm-thick epitaxial film of the magnetic MAX phase \MAX using time-resolved magneto-optical Kerr effect spectroscopy.
The demagnetization transients exhibit a two-step type-II process with a slow component time constant of $\sim$100~ps, which shows weak temperature and fluence dependence.
The first demagnetization stage is clearly present at high fluences but undetectable at low temperatures and low fluences.
Using the three-temperature model, we extracted the interaction constants and temperature-dependent spin heat capacity.
The spin-lattice coupling is comparable to that of other van der Waals magnets, while the electron-spin and spin-lattice couplings are several orders of magnitude smaller, consistent with the dominance of the slow demagnetization channel.
The reconstructed spin heat capacity exhibits weak temperature dependence, accounting for the absence of significant critical slowing down of demagnetization near $T_C$=220~K.
Our findings provide a starting point for experimental implementation of optical control of structures based on MAX phases, as proposed theoretically \cite{He-JPCL2020,Li-JPCL2022}, and introduce magnetic MAX phases and their derivatives magnetic MXenes \cite{JIANG_mag_MXene_review_2020} into the realm of optically-controlled 2D spintronics.

\section*{Acknowledgements}
The work of A.~A.~G and Ia.~A.~M. was supported by Russian Science Foundation (grant No.~24-72-00111).

\section*{Declaration of competing interest}
The authors declare that they have no known competing financial interests or personal relationships that could have appeared to influence the work reported in this paper.

\section*{Declaration of generative AI and AI-assisted technologies in the manuscript preparation process}
During the preparation of this work the authors used DeepSeek and Writefull to improve language and readability of the manuscript. After using this tool/service, the authors reviewed and edited the content as needed and take full responsibility for the content of the published article.

\printcredits
\bibliographystyle{elsarticle-num} 
\bibliography{cas-refs}

\end{document}